# DSMC investigation of rarefied gas flow through diverging micro- and nanochannels

Amin Ebrahimi[1], Ehsan Roohi[2]

High-Performance Computing (HPC) Laboratory, Department of Mechanical Engineering, Faculty of Engineering, Ferdowsi University of Mashhad, Mashhad, P.O. Box 91775-1111, Iran.

**Abstract**

Direct simulation Monte Carlo (DSMC) method with simplified Bernoulli-trials (SBT) collision scheme has been used to study the rarefied pressure-driven nitrogen flow through diverging microchannels. The fluid behaviours flowing between two plates with different divergence angles ranging between 0º to 17º are described at different pressure ratios ($1.5 \leq \Pi \leq 2.5$) and Knudsen numbers ($0.03 \leq Kn \leq 12.7$). The primary flow field properties, including pressure, velocity, and temperature, are presented for divergent microchannels and are compared with those of a microchannel with a uniform cross-section. The variations of the flow field properties in divergent microchannels, which are influenced by the area change, the channel pressure ratio and the rarefication are discussed. The results show no flow separation in divergent microchannels for all the range of simulation parameters studied in the present work. It has been found that a divergent channel can carry higher amounts of mass in comparison with an equivalent straight channel geometry. A correlation between the mass flow rate through microchannels, the divergence angle, the pressure ratio, and the Knudsen number has been suggested. The present numerical findings prove the occurrence of Knudsen minimum phenomenon in micro- and Nano- channels with non-uniform cross-sections.

**Keywords**: Divergent micro/Nanochannel; Rarefied gas flow; DSMC; simplified Bernoulli-trials; Knudsen minimum.

---

[1] - Current affiliation: Department of Materials Science & Engineering, Delft University of Technology, Mekelweg 2, 2628 CD Delft, The Netherlands.

[2] - Corresponding author; Associate professor; Email: e.roohi@um.ac.ir; Phone/fax: +98-513-8763304

# 1. Introduction

A deep understanding of fluid mechanics at the micro- and Nano-scales is important to get a better insight of microfluidic and nanofluidic processes, which can lead to the development of more efficient micro/Nano-electro-mechanical-systems (MEMS/NEMS). Channels play an essential role in microfluidic and nanofluidic devices in medical instruments, electronics cooling applications, laser equipment and automotive and aerospace industries. Because of the vast practical and theoretical importance of gaseous flows at the micro- and Nano-scales, many scholars have concentrated on this field of study as reviewed by Agrawal [1].

A decrease in the gas pressure and/or the length scales reduces the molecular collision rate and can make the continuum fluid assumption invalid. The degree of rarefication is generally determined by the Knudsen number (Kn) that is the ratio between the molecular mean free path ($\lambda$) and a characteristic length of the geometry. Based on the Knudsen number, four different rarefication regimes, including continuum ($Kn<10^{-3}$), slip ($10^{-3}<Kn<10^{-1}$), transition ($10^{-1}<Kn<10$), and free molecular ($Kn>10$), have been defined. It should be noted that this classification is based on the isothermal gas flow in long one-dimensional smooth channels and these ranges may vary for more complex flows. Micro- and Nano-devices usually operate in slip and transition regimes [2]. Different methods have been developed to numerically simulate the gaseous flows in microchannels. The direct simulation Monte Carlo (DSMC) method is one of the most common numerical techniques that has been widely employed for modelling the gaseous flows in various rarefication regimes [3]. DSMC is a probabilistic particle-based method to solve the Boltzmann equation based on the kinetic theory [4].

A change in the cross-section of flow passage may exist in some parts of the micro- or Nano-systems. Therefore, understanding the gas flow through such small geometries is vital for designing micro- and Nano-devices. Sharipov and Bertoldo [5] proposed a method for predicting the mass flow rate of rarefied gaseous flows in a long conical tube. Recently, this approach has been extended for long rectangular channels [6-8]. Stevanovic [9, 10] theoretically investigated the low-Mach number, isothermal and compressible gas flow in a two-dimensional microchannel of a variable cross-section by solving the Burnett momentum equation using the perturbation analysis. Veltzke *et al.* [11] presented an analytical model based on the superposition of the Poiseuille's law and a term that represents the molecular spatial diffusion to predict the mass flux of moderately rarefied gas flows ($Kn \approx 0.4$) in slightly tapered ducts. Hemdari *et al.* [12] experimentally investigated the occurrence of "Knudsen minimum" in divergent microchannels with different divergence angles up to 12°. They examined three different gases with different molecular weights and reported that the Knudsen minimum occurs in divergent microchannels.

Akbari *et al.* [13] developed a general predictive model to calculate the pressure loss in microchannels with an arbitrary cross-section shape and gradual variations in the cross-section under the continuum and slip regimes. Titarev *et al.* [14] numerically studied the gas discharge into vacuum through a divergent pipe of finite-length by solving the S-model kinetic equation. They reported that increasing the diameter of the pipe outlet leads to highly nonlinear flow patterns inside the pipe. Varade *et al.* [15] noticed the absence of the flow separation in their experiments of gas flow inside a suddenly expanding tube and observed a discontinuity in pressure distribution near the sudden expansion under the slip regime. Their findings show a qualitative similarity to the available two-dimensional numerical results [16]. Varade *et al.* [17] performed an investigation on nitrogen flow through divergent microchannels in the slip flow regime and the inlet-to-outlet pressure ratios higher than 9.5. They indicated that the viscous force is the dominant factor in the overall pressure loss of the gaseous flows in divergent microchannels. Duryodhan *et al.* [18] demonstrated that the flow reversal happens for liquid flows inside the divergent microchannels with a divergence angle larger than 16°, in contrast to the gaseous flows in the slip regime. The microscale gaseous flows are affected by the rarefaction effects and the augmented importance of gas-surface interaction, which could not be predicted accurately by the continuum theory. Duryodhan *et al.* [19] proposed a method for maintaining constant wall temperature condition in microdevices using divergent microchannels with divergence angles up to 8°.

An inspection of the available literature reveals that there is a lack of detailed understanding of gas flows in divergent micro- and Nano-channels. The primary focus of the previous studies on micro- and Nanochannel with sudden or gradual expansion has been devoted to continuum and slip rarefication regimes. A systematic and comprehensive study of the gas flow through divergent microchannels has been conducted with an interest in transitional rarefication regime, where the Navier-Stokes equations are not capable of predicting the gas flow behaviour accurately, and the results are presented in this paper. The primary aim of the present study is to scrutinise the effects of rarefication, cross-sectional area change and inlet-to-outlet pressure ratio on the gas flow behaviour in gradually diverging microchannels. The mass flow rate through divergent microchannels has been studied for different flow conditions and a correlation is suggested to estimate the mass flux in divergent microchannels. Additionally, the occurrence of the Knudsen minimum phenomenon in divergent microchannels has been investigated. The DSMC method with a novel collision scheme has been employed to reduce the computational costs.

## 2. Model description

The steady-state Poiseuille flow of a diatomic gas in divergent microchannels is investigated in this study. Figure 1 shows a schematic diagram of the model and the relevant parameters. The gas flow is described in a Cartesian coordinate system in which stream-wise direction is along the *x*-axis. The length of the channel (*L*) equals $20H_{in}$, where the height of the channel inlet ($H_{in}$) is 400 nm ($400 \times 10^{-9}$ m). Due to the symmetric flow field, only one-half the channel, the hatched region in Figure 1, was used for the computations. The Knudsen number was calculated based on the height of the channel inlet (*i.e.* Kn=$\lambda/H_{in}$). The divergence angle of the channel (*β*) was measured by the tangent of the angle between two solid walls (*i.e.* $\beta=(H_{out}-H_{in})/L$). Six channels with different values of *β* ranging from 0.0 to 0.3 were examined. The effect of cross-sectional area change on the flow behaviour is investigated by comparing the flow pattern in a microchannel of a uniform cross-section (*β*=0.0) with a non-uniform cross-section (*β*>0.0).

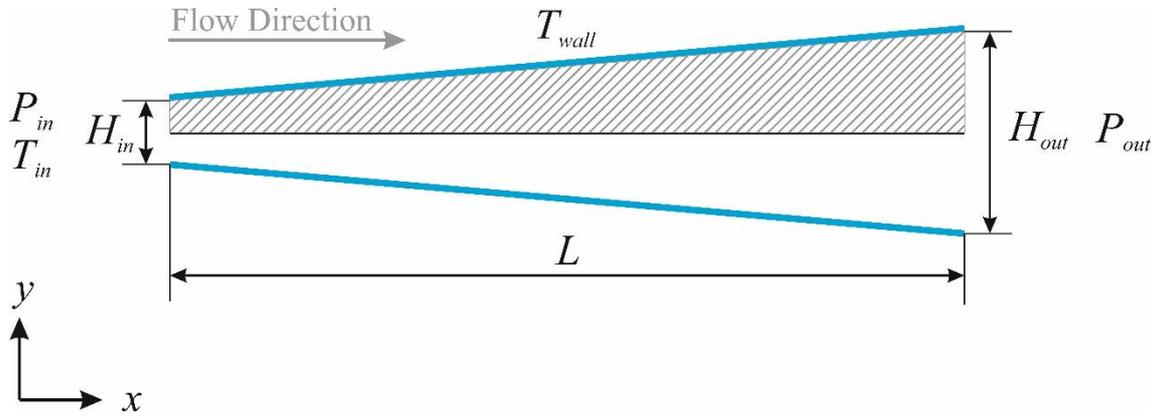

Figure 1- Schematic of the divergent microchannel and the associated parameters.

The inlet gas temperature ($T_{in}$) and the temperature of the solid walls ($T_{wall}$) were kept at 300 K. All solid walls were deemed to treat as fully diffuse, in another word, the tangential momentum and energy accommodation coefficient were considered as one. The inlet gas pressure ($P_{in}$) was prescribed based on the corresponding Knudsen number and the ideal gas law. Consequently, the outlet pressure could be determined using the specified pressure ratio ($\Pi$). The Knudsen number varies over a wide range of the rarefication regimes (0.03≤Kn≤12.7), including slip, transition and early-free-molecular regimes. The variable hard sphere (VHS) and Larsen-Borgnakke models were used for modelling the binary molecular collisions [4]. Nitrogen was selected as the working fluid and the energy exchange between both rotational and translational modes and the energy transfer from and to the internal modes were allowed. The molecular properties of nitrogen are summarised in Table 1, where, $d_p$, $m_p$, $\omega$ and $DOF_{rot}$ stand for molecular diameter, molecular mass, viscosity index and rotational degrees of freedom of the gas molecules, respectively.

Table 1- Molecular properties of nitrogen molecules.

| $d_p$ [m] | $m_p$ [kg] | $\omega$ [-] | $DOF_{rot}$ [-] |
|---|---|---|---|
| 4.17×10$^{-10}$ | 4.650×10$^{-26}$ | 0.74 | 2 |

## 3. Numerical procedure

The DSMC method [3, 4] is a computational technique for modelling a dilute gas in which the molecular diameter is much smaller than the molecular mean free path. A number of simulator particles are used in this method to represent any number of real molecules or atoms, and the particle collisions and movements during a specified time step smaller than the local mean collision time are decoupled. Each of these discrete simulator particles has their own position and velocity, and their states are updated and stored at every time step. Particle ballistic movements in the physical space are modelled deterministically, and the inter-molecular collisions and the particle-surface interactions are handled probabilistically. Macroscopic properties of the flow field in each computational cell can be obtained by spatial and time averaging of the microscopic particle properties [4]. The DSMC method is a robust numerical approach for modelling the gaseous flows, especially in the transitional rarefication regime.

A validated open-source DSMC solver, dsmcFOAM, which is available within the framework of the OpenFOAM [20], is employed in the present study. The solver has been equipped with a set of subsonic, implicit pressure boundary conditions [21, 22] and the accuracy and reliability of the solver have been meticulously verified against the experimental and theoretical data [22-25]. It is worth mentioning that this kind of boundary conditions at the channel inlet and outlet implicitly assumes an equilibrium flow up to these boundaries. The "Simplified Bernoulli Trials (SBT)" [26] collision partner selection model was implemented within the solver. The SBT collision scheme has been introduced as an alternative to traditional collision schemes, such as no time counter (NTC) [4], to reduce the statistical fluctuations and the computational costs by avoiding repeated collisions. The main advantage of the SBT collision scheme is the capability of predicting the flow field accurately with a smaller number of particles per cell (PPC), which results in less need of computational resources. One can find more detailed information about the SBT collision scheme algorithm and its applications for modelling various flow fields in [26-29]. All simulations were executed in parallel on four cores of an Intel Core i7-3770K (up to 3.90 GHz) processor. Upon the total population of simulator particles and the average linear kinetic energy of the system reached almost constant values or the mass flow rates at the inlet and the outlet became equal, the solution was presumed to be steady. Subsequently, the simulation was continued and time averaging was performed by $2.5 \times 10^6$ times sampling of the instantaneous DSMC particle fields to reduce the statistical scatter and obtain a smooth flow field after achieving the steady-state condition.

The number of particles per cell, the cell size and the time step can affect the accuracy and reliability of a DSMC simulation. Since the DSMC solver equipped with the SBT collision scheme can provide accurate results with two or even less than two simulator particles per computational cell [26], all simulations were initialised with PPC=5 to have a sufficient number of particles near the outlet. However, a PPC independence test was accomplished to make the statistical relationships between particles negligible and prevent costly computations. It has been demonstrated that employing very fine grid sizes in the stream-wise direction is not essential for modelling the gas flows in microchannels due to the lack of high flow gradients [30]. The grid independence study was performed to select an appropriate grid for implementing reliable intermolecular collisions, providing a convenient physical space for sampling the macroscopic properties and avoiding high computational costs. A structured mesh with cell sizes of $\Delta x = \lambda_{in}$ and $\Delta y = \lambda_{in}/3$ in the streamwise and the normal directions, respectively, was chosen where $\lambda_{in}$ is the molecular mean free path at the channel inlet. The computational grid for the cases with inlet Knudsen numbers ($Kn_{in}$) higher than 0.25 was chosen same as the cases with $Kn_{in}=0.25$ to capture the flow gradients. The time step size was determined smaller than the local mean collision time to allow successful decoupling of the collision and movement procedures and sufficiently small that simulator particles are likely to stay multiple time steps in a single cell. Our past experiences on numerical modelling of fluid flow and heat transfer in micro- and Nano-channels [17, 31-34] are extended here.

## 4. Results and discussions

The variations of normalised static pressure along the channel centreline are shown in Figure 2 for different divergence angles (0.0<$\beta$<0.3) and three different rarefication regimes. The pressure distribution is non-linear for compressible flows in microchannels of uniform cross-section and tends to a linear distribution with increasing the rarefication effects [35]. The non-linearity of the static pressure distribution in a channel of uniform cross-section is due to the compressibility of the flow and hence is more pronounced at higher pressure ratios. An increase in the divergence angle attenuates the compressibility effects and intensifies the effects of gas expansion, *i.e.* it reduces the local molecular population, which can result in a change in the concavity of pressure distribution curve. It is seen that the curvature of the pressure distribution curve decreases with increasing the Knudsen number. A comparison of the molecular population distribution in the divergent microchannels with the microchannels of uniform cross-section using the same boundary conditions implies that increasing the divergence angle leads to a sharper density reduction in the microchannel entrance region. This change in density distribution can change the concavity of the pressure distribution curve. For the case with $\beta$=0.3 and $Kn_{in}$=0.03, the pressure gradients are stronger at x/L<0.2 and the pressure drops below the outlet pressure around x/L=0.4.

It is obvious that an orifice flow will form on the upper limit of *$\beta$*. It can be argued that the effect of increasing the divergence angle of the channel on the pressure distribution along the channel is analogous to a reduction of the length-to-height ratio (*L/H*) of the equivalent channel of uniform cross-section with a finite length. Accordingly, an increase in the divergence angle of the channel and hence the local Knudsen number inside the channel makes the locus of the maximum deviation from linear distribution closer to the channel inlet. This argument is consistent with the concept of equivalent hydraulic diameter proposed in Ref. [17, 18] and the results provided in Ref. [36].

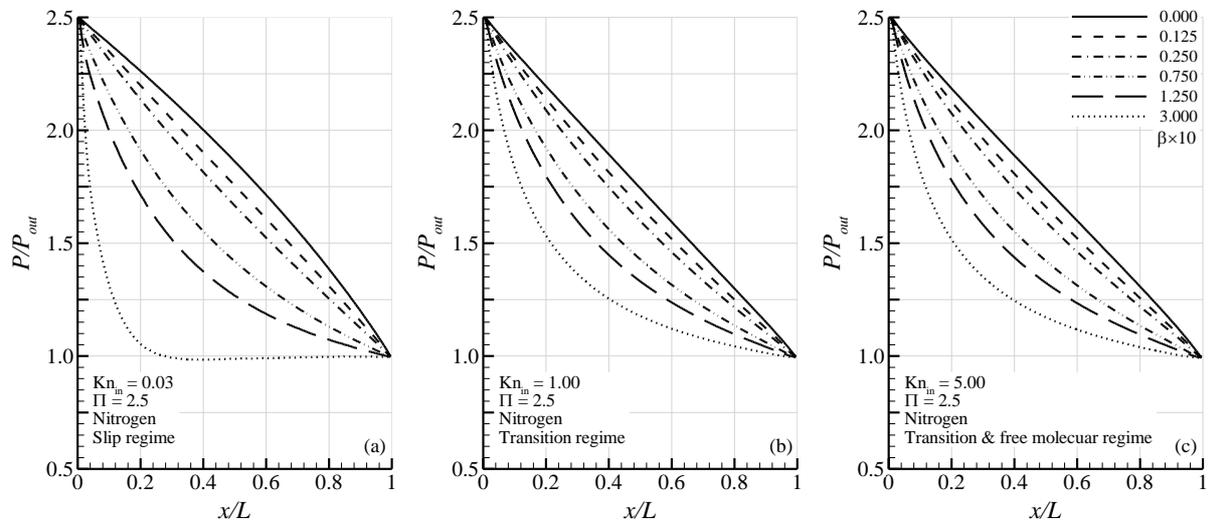

Figure 2- Static pressure profiles (normalised with outlet static pressure) along the centreline of diverging channels with different divergence angles and Π=2.5. (a) Slip regime; (b) Transition regime; (c) Transition and early-free molecular regime.

Distribution of the centreline Mach number (Ma) is shown in Figure 3. Higher gas velocities have been found in the divergent microchannels compared with the microchannels of uniform cross-section, which boost up with an increase in the divergence angle. In a microchannel with a uniform cross-section, the gas flow is dominated by the imposed pressure gradients, and the velocity increases along the channel. It is seen that the Mach number distribution along the microchannels changes from an increasing to a decreasing distribution with increasing the divergence angle. A slight increase in the divergence angle (up to β=0.075) decreases the rate of the fluid acceleration along the microchannel, although the driving pressure force still dominates the decelerating effects of area increase and friction force and consequently no deceleration is seen in the cases with β<0.075. The driving pressure force is balanced with the decelerating effects of area increase and friction force for the gas flows in divergent microchannels with β=0.075. It can be argued that for any inlet-to-outlet pressure ratio there is a specific divergence angle that results in a constant gas velocity along the channel. A further increase in the divergence angle augments the area increase effect, which results in subsonic fluid deceleration along the channel.

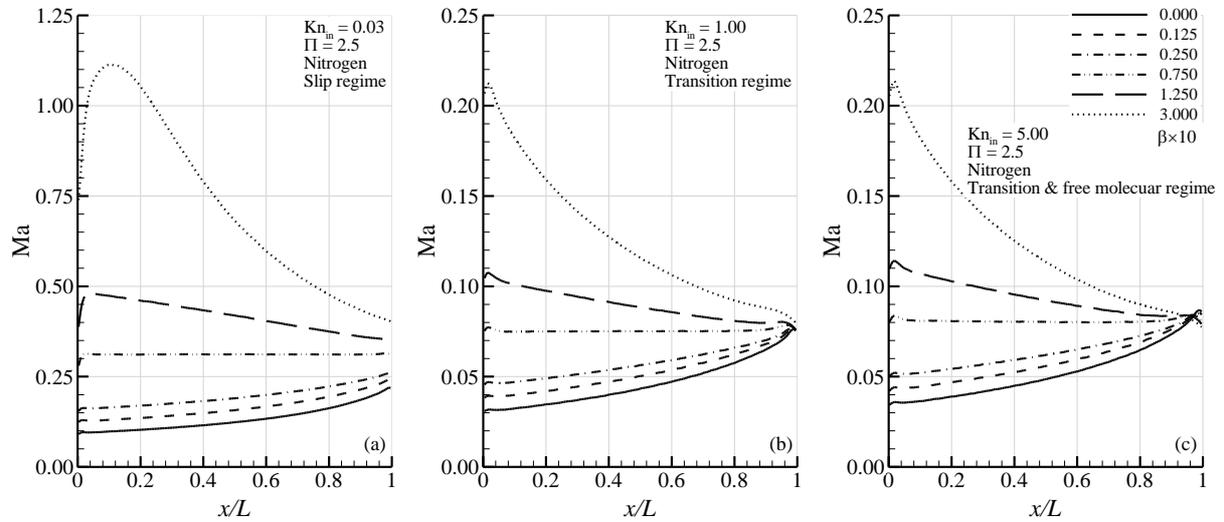

Figure 3- Mach number distributions along the centreline of diverging channels with different divergence angles and Π=2.5. (a) Slip regime; (b) Transition regime; (c) Transition and early-free molecular regime.

An increase in the fluid velocity is observed in the entrance region of the microchannels with large divergence angles ($\beta \geq 0.075$). The fluid accelerates rapidly and reaches a supersonic speed in the case with $\beta$=0.3 and $Kn_{in}$=0.03 due to the higher pressure drop near the channel inlet, see figure 2-a. This observation can be described by looking at the viscous boundary layer development. For the channels with large divergence angles, the interaction between the thick laminar boundary layer growth and the actual diverging geometry introduces a virtual throat, which is responsible for the fluid acceleration from a subsonic speed at the inlet of a purely diverging passage. Figure 4 illustrates the contours of Mach number, streamlines, velocity vectors and the axial and transversal density gradients in a microchannel with $\beta$=0.3, $Kn_{in}$=0.03 and Π=2.5. In this case, the fluid reaches a supersonic speed passing through the virtual throat depicted in figure 4-a. The virtual throat could also be displayed by the transversal density gradients shown in figure 4-d. Figure 5 represents the variations of Mach number and axial density gradients along the channel centreline for the case with $\beta$=0.3, $Kn_{in}$=0.03 and Π=2.5. It is seen that the fluid decelerates moving towards the channel outlet due to the strong viscous effects to adapt the outlet pressure. None of the numerical Schlieren images shown in figure 4 d and e indicates the occurrence of a shock wave. Therefore, the weak supersonic flow reverts to the subsonic regime and the Mach number decreases due to the strong diffusion effects [37].

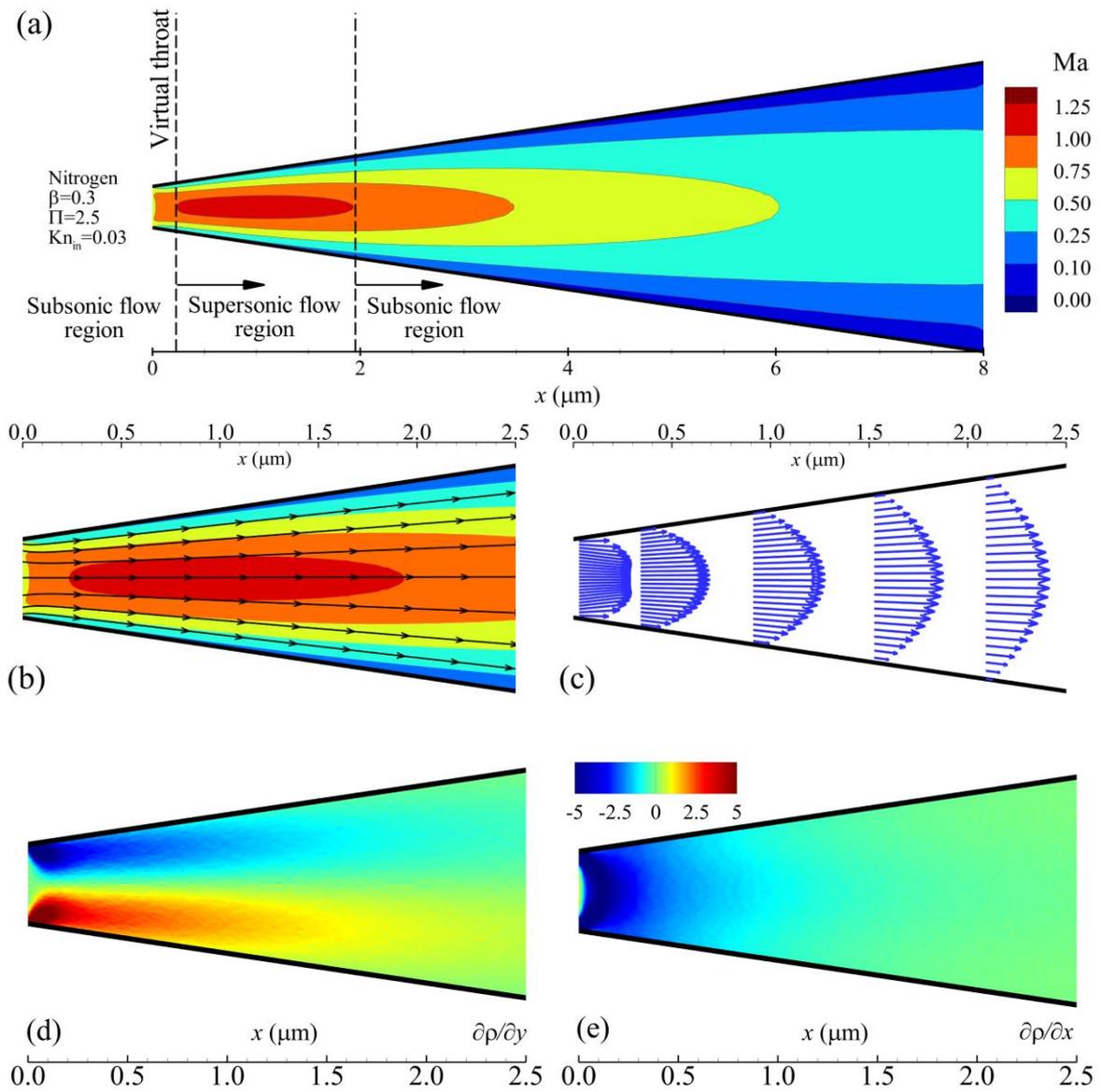

Figure 4- (a) Contours of Mach number, magnified view of (b) streamlines, (c) velocity vectors at different sections, (d) transversal density gradients $\partial\rho/\partial y$, and (e) axial density gradients $\partial\rho/\partial x$ of a diverging microchannel with $\Pi=2.5$, $Kn_{in}=0.03$, and $\beta=0.3$.

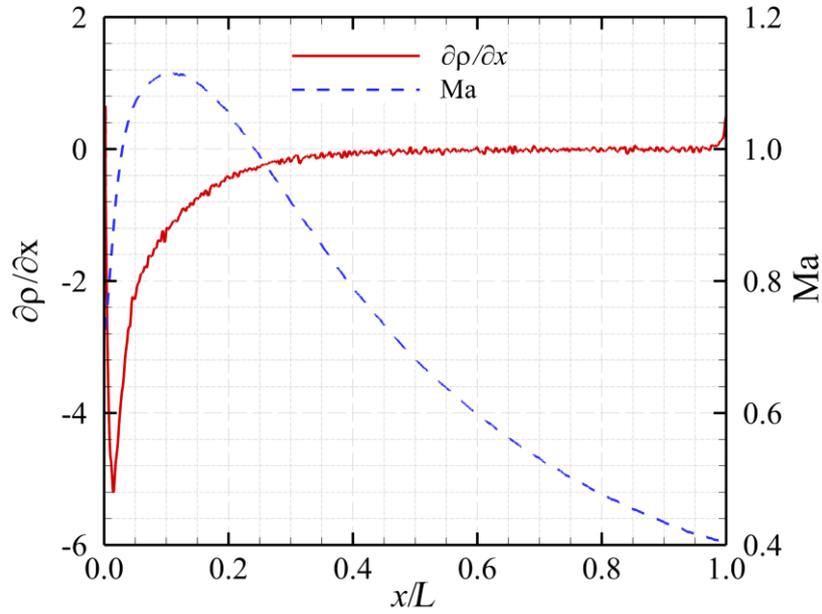

Figure 5- Variations of Mach number and axial density gradients ($\partial\rho/\partial x$) along the channel centreline for the case with $\beta$=0.3, $Kn_{in}$=0.03 and $\Pi$=2.5.

Figure 6 shows the variations of the slip Mach number on the microchannel wall for different cases with $\Pi$=2.5. The slip-velocity has been defined as the velocity difference between the wall and the gas adjacent to the wall. It is found that the larger the divergence angle of the channel, the higher the slip velocity on the channel wall. A similar pattern is expected for the shear stress distribution because of the direct relationship between the slip velocity and the shear stress, reads from Maxwell's velocity slip relation [3]. No flow separation was observed in the divergent microchannels for the range of parameters studied in the present work. It should be remarked that the absence of flow separation could also be deduced from the Figure 2 where no adverse pressure gradients was observed for most of the cases. The absence of flow reversal can be attributed to the strong viscous effects in the low Reynolds number flows. Since the driving pressure force of the rarefied Poiseuille flow in micro- and Nano-channels is much higher than its kinetic energy [15, 38], a change in the kinetic energy due to the cross-sectional area change has not a significant effect on the pressure gradients. Additionally, the higher momentum diffusion in rarefied gas flows compared with the continuum

Page **12** of **25**

flows [15, 39] makes the gas molecules to follow the surface pattern. The slip on the channel walls is another reason for keeping the flow separation at bay.

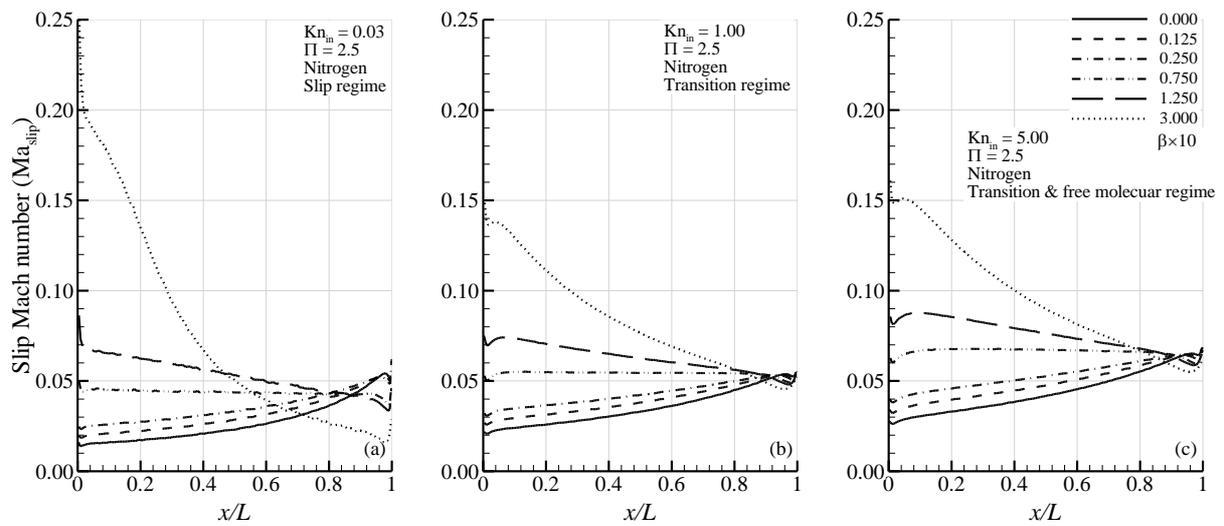

Figure 6- Slip velocity distributions along the surfaces of diverging channels with different divergence angles and $\Pi=2.5$. (a) Slip regime; (b) Transition regime; (c) Transition and early-free molecular regime.

A shift in the slip velocity is observed by increasing $Kn_{in}$. To study the effect of Knudsen number on the flow field, the averaged Mach number on the channel centreline and along the channel wall (slip Mach number) is presented in Figure 7 for different Knudsen numbers. It is clearly seen that the magnitude of the averaged slip Mach number increases with increasing the Knudsen number. This occurs due to the intensification of the non-equilibrium effects. The rate of the collisions between the gas molecules and the solid walls reduces under the rarefied condition that results in a slip velocity by statistical averaging [3] and introduces the non-equilibrium effects. The averaged centreline Mach number decreases with increasing the Knudsen number from 0.03 to 1 then it slightly increases with a further increase in the Knudsen number. The effect of the slip velocity on the centreline velocity increases with Knudsen number due to the higher momentum diffusion at high Knudsen numbers. This increase in the centreline Mach number is more noticeable for the channels with low divergence angles since any increase in the cross-sectional area diminishes the effects of molecular diffusion.

The rate of effective gas viscosity reduction decreases for Knudsen numbers greater than one [40] and the driving pressure gradients can accelerate the fluid in the channel at $Kn_{in}>1$.

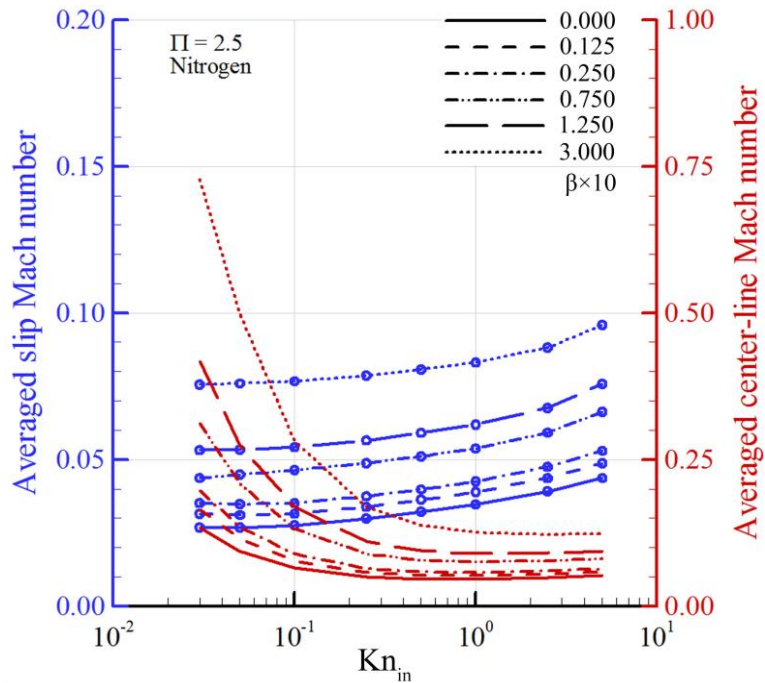

Figure 7- Effects of Knudsen number on the averaged centreline and slip Mach number. ($\Pi$=2.5; red line: averaged centreline Mach number, blue line with symbols: averaged slip Mach number)

Based on the conservation of energy, an increase in the kinetic energy of the gaseous flows leads to a decrease in the fluid internal energy and hence the fluid temperature. The variations of fluid temperature along the channel centreline are shown in Figure 8 for different divergence angles. It is expected that the heat transfer rate reduces with increasing the Knudsen number, because of the lower molecular collision rate, and intensifies with an increase in the inlet-to-outlet pressure ratio due to the higher molecular collision frequency. A temperature drop was observed near the channel outlet for the subsonic cases, which can be attributed to the rapid gas expansion near the outlet. The higher the inlet-to-outlet pressure ratio, the higher the fluid velocity and the larger the temperature drop. A large temperature drop occurs in the case with $\beta$=0.3 and $Kn_{in}$=0.03 due to the rapid fluid acceleration downstream to the channel inlet. The fluid not only decelerates flowing through the highly diverging channels but also the fluid temperature increases due to the kinetic energy conversion into the internal energy. It can be argued that for the case with *β*=0.3 and Kn$_{in}$=0.03, the stronger fluid deceleration rate leads to an augmented kinetic energy conversion into the internal energy, which results in the higher fluid temperature at the outlet.

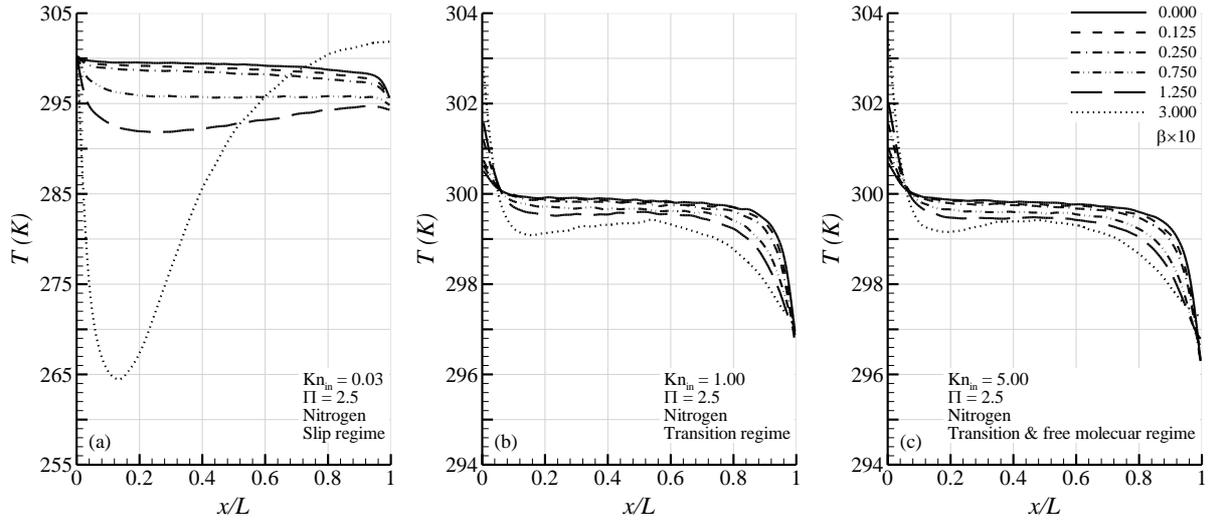

Figure 8- Temperature distributions along the centreline of diverging channels with different divergence angles and Π=2.5. (a) Slip regime; (b) Transition regime; (c) Transition and early-free molecular regime.

The mass flux through the divergent microchannels has been normalised by the corresponding mass flux through the microchannel with a uniform cross section (*i.e.* $j = \dot{M}/\dot{M}_{straight}$) and the results are presented in Figure 9. It was spotted that for the range of parameters studied here, the amount of the mass that a divergent microchannel can carry increases with the divergence angle and is higher than a microchannel with a uniform cross-section. However, the magnitude of *j* declines with increasing the Knudsen number. The increase in the mass flux with increasing the divergence angle may be attributed to the increase of the pressure force to the friction force ratio in the diverging channels due to the weakened wall effects. Similar behaviour has been already reported from the experimental observations of the rarefied gas flow through short tubes with changing the length-to-radius ratio [41]. This observation reinforces the argument that, in some aspects (pressure distribution and mass flow rate), the influence of increasing the divergence angle of the channel is analogous to

Page **15** of 25

the reduction of the length-to-height ratio (*L/H*) of the equivalent channel of uniform cross-section with a finite length. The mass flux ratio is almost identical for different pressure ratios, evidently in the transition regime. The intensified non-equilibrium effects at the Knudsen numbers higher than 1 results in a change in the slope of the mass flow rate variations with the Knudsen number. For any fixed divergence angle and Knudsen number, there is a critical pressure ratio that further reduction in the back pressure to increase the pressure ratio does not result in the mass flow rate increase [31, 42]. Decreasing the Knudsen number increases this critical pressure ratio [43]. Based on the results of the present study, the following correlation is suggested for estimating the mass flow rate ($\dot{M}$) through diverging microchannels.

$$\dot{M} = \frac{a(0.1+\beta)^{1.678}}{Kn_{in}^{1.675}} \Pi \qquad (0.0 \leq \beta \leq 0.3; 0.03 \leq Kn_{in} \leq 5.0; 1.5 \leq \Pi \leq 2.5) \qquad (1)$$

where *a* is a constant multiplier equals to $3.238 \times 10^{-6}$ [Kg s$^{-1}$]. The proposed correlation fits the DSMC data points with an R-square of 0.9985 and a root mean squared error (RMSE) of $1.61 \times 10^{-6}$%. Figure 10 shows a comparison between the predicted mass flux from Eq. 1 and the DSMC results at various Knudsen numbers and pressure ratios.

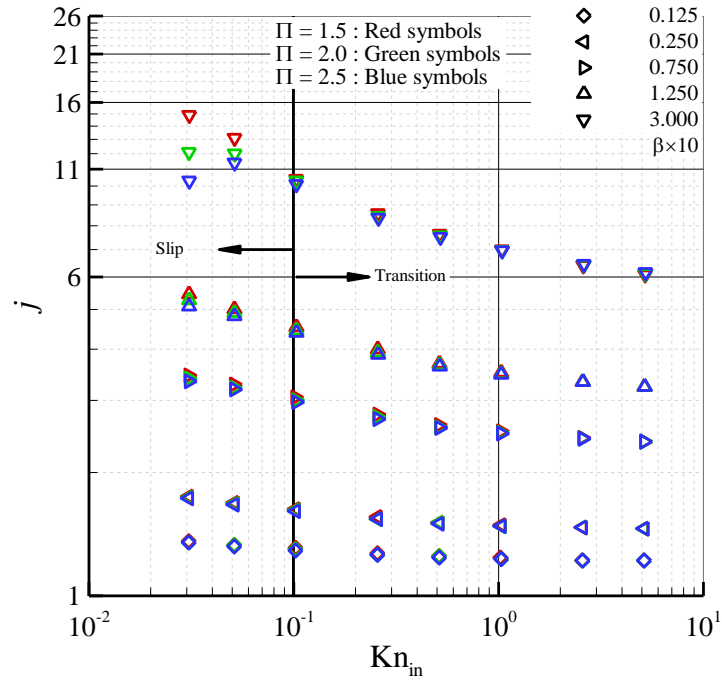

Figure 9- Variations of the mass flux in diverging microchannels (normalised with the mass flux of the corresponding straight microchannel) with inlet Knudsen number.

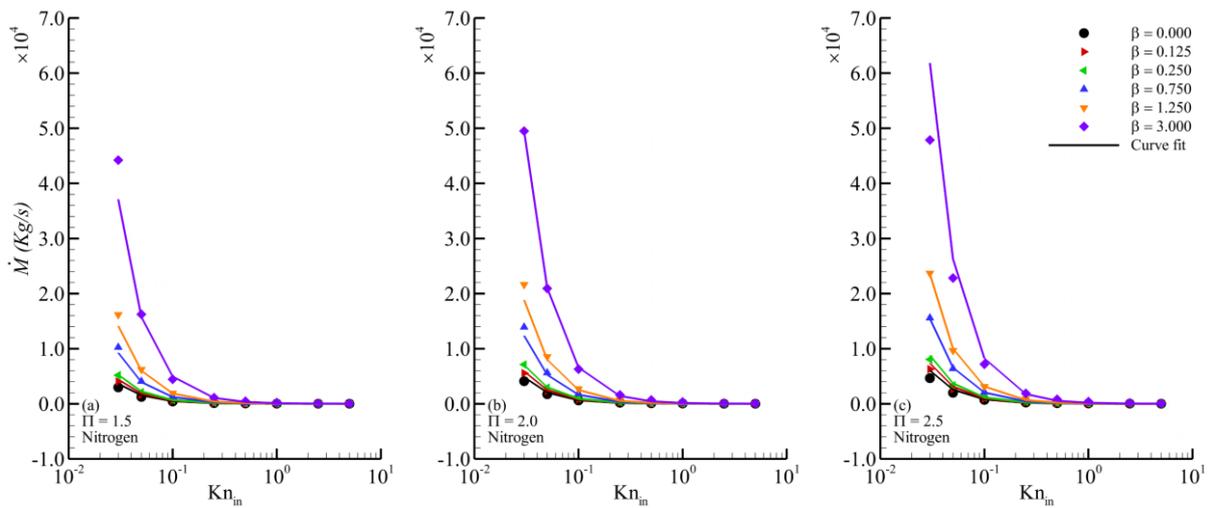

Figure 10- Comparison of the predicted mass flux in diverging microchannels (: Line) with obtained DSMC results (: Symbol) at various Knudsen numbers for (a) $\Pi = 1.5$; (b) $\Pi = 2.0$; (c) $\Pi = 2.5$.

Knudsen [44] and Gaede [45] reported a mass flow rate minimum for the channel and pipe flows in the transition regime at Kn≈1. This phenomenon, called "Knudsen paradox" or "Knudsen minimum", is well known in the rarefied micro- and Nano-fluidics field of studies. The dimensionless mass flow rate is presented in Figure 11 to investigate the Knudsen minimum phenomenon in the divergent microchannels. In Figure 11, $Kn_m$ represents the Knudsen number at the arithmetic mean Knudsen number of the inlet and the outlet. The normalised mass flow rate ($M_n$) was calculated using Eq. 2 [46]:

$$M_n = \frac{L\sqrt{2RT}}{(P_{in} - P_{out})H_{in}^2 w} \dot{M} \qquad (2)$$

where $R$ is the specific gas constant [J kg$^{-1}$ K$^{-1}$], $\dot{M}$ is the mass flow rate [kg s$^{-1}$] obtained from the DSMC simulations and $w$ is the channel depth [m] and was assumed to be 1.0 for the two-dimensional cases. Figure 11 illustrates the existence of the Knudsen minimum phenomenon in the case of diverging microchannels, however, this minimum becomes unclear with an increase in the divergence angle. It is seen that the Knudsen minimum occurs at $Kn_m \approx 1$ and the minimum moves towards higher Knudsen numbers with increasing the divergence angle and the inlet-to-outlet pressure ratio. The occurrence of this phenomenon has been attributed to the larger contribution of the diffusive mass flux with respect to the convective mass flux in the total mass flow rate through the channel. Since the Knudsen number is inversely proportional to the Reynolds number (*i.e.* $Kn \propto Re^{-1}$), an increase in the Knudsen number leads to an increase in the friction factor results in a reduction in the convection. It has been found that the variations of the gas density are small for the Knudsen numbers greater than 1 and the diffusive transport process starts to be independent of density and be proportional to the pressure gradients. Referring to Figure 7, increasing the Knudsen number initially causes a rapid reduction in the centreline velocity and a slight increase in the slip velocity and then, at a certain Knudsen number, the rate of the slip velocity increase becomes greater than the rate of the centreline velocity reduction. The occurrence of the minimum in normalised mass flow rate can be justified by considering the variations of the density, slip velocity, and centreline velocities with the Knudsen number.

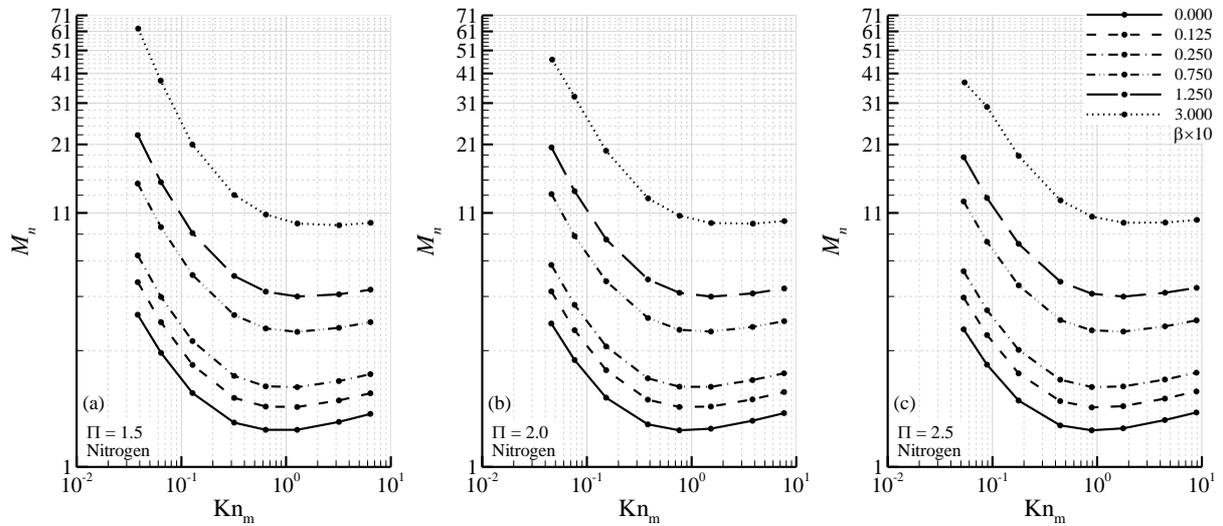

Figure 11- Dimensionless mass flow rate through diverging microchannels versus mean Knudsen number for (a) Π=1.5; (b) Π=2.0; (c) Π=2.5.

## 5. Conclusion

The rarefied gas flows through the divergent microchannels were numerically studied using the DSMC method equipped with the SBT collision scheme. The fluid behaviour was described for gas flows in divergent microchannels with various divergence angles and pressure ratios under different rarefication regimes. It was explained that an increase in the divergence angle intensifies the ratio of the pressure force to the friction effects, which leads to an increase in the mean gas velocity. It was argued that for any inlet-to-outlet pressure ratios there is a specific divergence angle that results in a constant fluid velocity along the channel, in which the pressure force and the friction force are balanced with the decelerating effect of the area increase. No flow separation was observed in the divergent microchannels studied in the present paper due to the slip on the walls and the higher viscous diffusion under rarefied conditions. An augmentation in the mass flux was found in divergent microchannels compared with the microchannels of uniform cross-section, which is more intensified at lower Knudsen numbers. The Knudsen minimum phenomenon was spotted in the microchannels with a diverging cross-section. This phenomenon occurs at higher Knudsen numbers with increasing the divergence angle and the inlet-to-outlet pressure ratio. Additionally, a correlation between the mass flow rate through microchannels, the divergence angle, the pressure ratio, and the Knudsen number was suggested.

**Acknowledgements**

The authors would like to thank Prof. Ali Beskok from the Southern Methodist University, the USA, and Prof. Yevgeny A. Bondar from the Novosibirsk State University, Russia for the fruitful discussions concerning the results presented in this paper.

**Nomenclature**

| | |
|---|---|
| *a* | Constant [kg s$^{-1}$] |
| DOF$_{rot}$ | Rotational degree of freedom |
| $d_p$ | Molecular diameter [m] |
| DSMC | Direct simulation Monte-Carlo |
| *H* | Channel height [m] |
| Kn | Knudsen number |
| *L* | Channel length [m] |
| Ma | Mach number |
| $\dot{M}$ | Mass flow rate [kg s$^{-1}$] |
| $m_p$ | Molecular mass [kg] |
| *n* | Number density [m$^{-3}$] |
| NTC | No time counter |
| *P* | Pressure [Pa] |
| PPC | Particle per cell |
| *R* | Specific gas constant [J kg$^{-1}$ K$^{-1}$] |
| RMSE | Root mean squared error |
| SBT | Simplified Bernoulli-trials |
| *T* | Temperature [K] |
| VHS | Variable hard sphere |
| *w* | Channel depth [m] |
| *x*, *y* | Cartesian coordinates |

Greek symbols

| | |
|---|---|
| $\beta$ | Divergence angle |
| $\kappa_b$ | Boltzmann constant [m$^2$ kg s$^{-2}$ K$^{-1}$] |
| $\lambda$ | Molecular mean free path [m] |
| $\Pi$ | Inlet-to-outlet pressure ratio |
| $\rho$ | Density [kg m$^{-3}$] |
| $\omega$ | Viscosity index |

Subscripts

| | |
|---|---|
| in | Inlet |
| m | Mean |
| n | Normalised |
| out | Outlet |
| slip | Slip |
| straight | Straight |